\documentclass[a4paper, 11pt]{article}

\usepackage{a4wide}

\usepackage[english]{babel}
\usepackage[utf8]{inputenc}
\usepackage[T1]{fontenc}

\usepackage{amsthm}

\usepackage[ttscale=.875]{libertine}
\usepackage[libertine]{newtxmath}
\usepackage[scaled=1]{inconsolata}

\usepackage{graphicx}

\theoremstyle{remark}
\newtheorem*{remark}{Remark}
\newtheorem{idea}{Idea}

\usepackage{booktabs}

\usepackage[colorlinks=false, hidelinks]{hyperref}

\title{Revisiting Monte Carlo Strength Evaluation}
\author{
  Martin Stanek \\[2ex]
  Department of Computer Science \\
  Faculty of Mathematics, Physics and Informatics \\
  Comenius University \\
  \textsl{martin.stanek@fmph.uniba.sk}
}

\date{}

\begin{document}
\maketitle

\begin{abstract}
The Monte Carlo method, proposed by Dell'Amico and Filippone, estimates a password's rank within a probabilistic model for password generation, i.e., it determines the password's strength according to this model. We propose several ideas to improve the precision or speed of the estimation. Through experimental tests, we demonstrate that improved sampling can yield slightly better precision. Moreover, additional precomputation results in faster estimations with a modest increase in memory usage. \\[2ex]
\end{abstract}

\section{Introduction}

Passwords remain a frequently used authentication method, despite numerous initiatives, technologies, and implementations aiming for passwordless authentication. Although the popularity of methods such as Windows Hello, Passkey, and WebAuthn has increased, the security of passwords continues to be a significant topic in many application areas.

Evaluating the strength of a password is useful for providing users with feedback on their chosen passwords. This feedback can assist users in selecting stronger passwords. Often, the strength is calculated as a password's rank, i.e., how many passwords will be generated by some chosen algorithm until our password is produced. 
There are various tools that calculate the strength of the password, for example zxcvbn \cite{W16}, or password scorer tool in PCFG cracker \cite{pcfgcracker}.

Dell'Amico and Filippone proposed a Monte Carlo algorithm that estimates a password's rank within a probabilistic model \cite{AM15}. The algorithm work for any probabilistic password generation model, and the authors proved that estimated results converge to the actual ranks.

The Monte Carlo estimator is also used to evaluate and compare different probabilistic models for password generation. The original paper compares $n$-grams models \cite{N05}, the PCFG model using probabilistic context-free grammar \cite{W09}, and the Backoff model \cite{M14}. Recent example of using the Monte Carlo estimator is the evaluation of a password guessing method that employs a random forest \cite{WZZX23}.

\paragraph{Our contribution.} We propose three ideas for improving the precision or speed of the Monte Carlo estimator. The first idea is to interpolate password's rank within the sampled interval it belongs, according its probability. The second idea aims to reduce probability overlap in sampled passwords. Both these ideas, presented in Section \ref{sec-ideas12}, seek to improve the estimator's precision. The estimation speed for a password, originally based on binary search, can be enhanced with some additional data computed in advance (the third idea, see Section \ref{sec-idea3}). All ideas have been tested experimentally to assess their merit. The results are presented in Section \ref{sec4}.
Our experiments demonstrate that improved sampling can yield slightly better precision. However, the effect of interpolation on precision is inconclusive, and we cannot rely on this technique to improve precision.

We utilize the reference implementation of the Monte Carlo estimator, which was published by one of the authors of the original paper on GitHub \cite{A16}, and we employ the RockYou dataset for our experiments. Given that our focus lies on the estimator itself, the choice of dataset is relatively unimportant.

\section{How the Monte Carlo Estimator works}
\label{sec2}

We mostly follow \cite{AM15} in this section. Let $\Gamma$ be a set of all allowed passwords. A probabilistic password model aims to capture how humans select password, assigning higher probabilities to more frequently chosen passwords and lower probabilities to less common ones. Let $p(\alpha)$ denotes a probability assigned to password $\alpha$ by the model, such that $\sum_{\alpha\in \Gamma} p(\alpha)= 1$. Different models yield different probability distributions.

When the model is used for an attack, it enumerates password in descending order of probability. Therefore, the strength of a password $\alpha$ is the number of passwords with a higher probability: 
\[
  S_p(\alpha) = |\{ \beta\in\Gamma;\; p(\beta)>p(\alpha) \}|.
\]

\begin{remark} 
In this context, the authors do not address the possibility that the model may assign identical probabilities to multiple passwords, resulting in a non-monotonic
$p$. The definition of $S_p(\alpha)$ assigns all passwords that share the same probability the lowest rank in their group. This approach can be considered prudent from a security standpoint.
\end{remark}

Computing the exact value of $S_p(\alpha)$, for a random $\alpha$, has prohibitively large time complexity. The Monte Carlo estimator uses sampling and approximation to provide efficient and sufficiently accurate estimation. It relies on two properties of the underlying model:

\begin{itemize}
\item The model allows for efficiently computing $p(\alpha)$ for any password $\alpha$.
\item There is an efficient sampling method that generates a password according to the model's distribution.
\end{itemize}

\paragraph{Precomputation.}

The estimator generates a sample $\Theta$ of $n$ passwords (sampling with replacement). The sample $\Theta = \{\beta_1,\ldots, \beta_n\}$ is sorted by descending probability, i.e., $p(\beta_1)\geq\ldots\geq p(\beta_n)$. The cumulative ranks of sampled passwords are calculated as follows:
\[
  c_i = \frac 1n \, \sum_{j=1}^i \frac{1}{p(\beta_j)}\quad\text{ for }\; i=1,\ldots, n. 
\]

The estimator needs to store the probabilities. The cumulative ranks can be easily recomputed. However, both these arrays are usually significantly smaller than representation of the model, see Section \ref{sec-areas}.

\begin{remark} 
The implementation \cite{A16} uses negative $\log_2$ probabilities, i.e., scaling $p(\beta_j)$ to $-\log_2 p(\beta_j)$.
\end{remark} 

\paragraph{Estimation.}

In order to estimate $S_p(\alpha)$ for some password $\alpha$, the probability $p(\alpha)$ is computed first. Then the binary search is used to compute the largest  index $j$ such that $p(\beta_j) > p(\alpha)$. The result, estimated rank of $\alpha$ is $S_p(\alpha) \approx c_j$. Hence, the time complexity of the estimator is $O(\log n)$.

\section{Areas for improvement}
\label{sec-areas}

\subsection{Memory requirement}

\begin{figure}[h]
\centering
\includegraphics[scale=0.65]{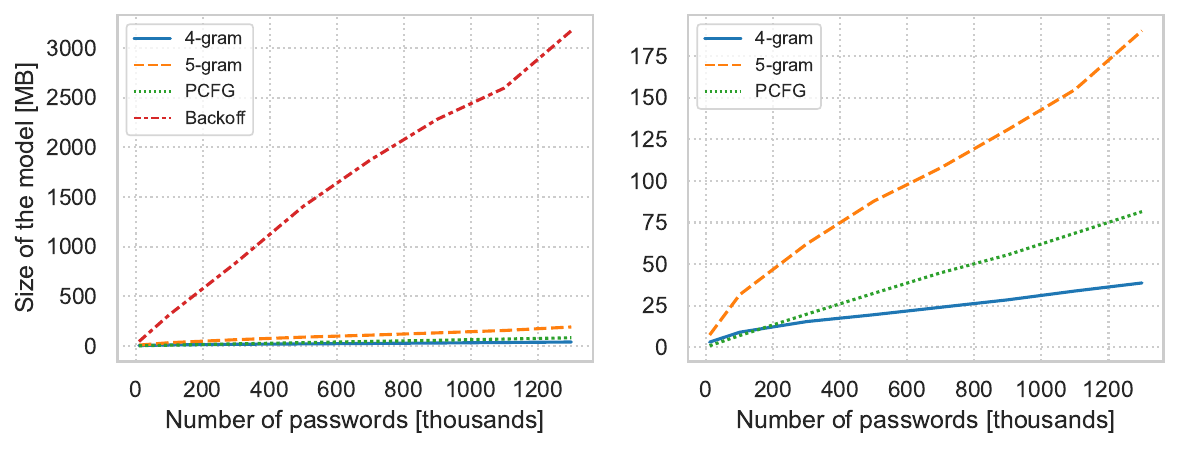}
\caption{Size of the model reflecting the number of passwords in a training dataset. The graph on the right excludes the Backoff model to show other three models more clearly.}
\label{fig1}
\end{figure}

The RockYou dataset contains more than 14 million unique passwords. The more passwords are used to train a model, the better and more precise results we can expect, such as in our case for password strength estimation. However, there is a point beyond which additional training data provide only negligible improvement, while further increasing the model's size. Notably, even the set of 10,000 most frequent passwords generates models of substantial size: $3.17\,$MB for 4-gram, $7.45\,$MB for 5-gram, $43.5\,$MB for Backoff, and $0.99\,$MB for PCFG. An attempt to use up to $10$\% of the RockYou dataset for training leads to unacceptable model sizes, where Backoff model being the largest, as shown in Figure \ref{fig1}.

The model defines how passwords are represented, generated, and how their probabilities are calculated. Since these methods are specific for each model, we do not aim to improve the model size. However, the Monte Carlo estimator utilizes an additional arrays, where probabilities and ranks of sampled passwords are precomputed. The original paper \cite{AM15} experiments with various sample sizes up to 100,000 (having ``\textsl{relative error $1$\%}''), but mostly uses the default sample size of 10,000. The default sample size requires 160\,kB of memory\footnote{Real numbers are represented as the \texttt{numpy.float64} datatype.} and its dominated by the memory required for any model trained on a dataset of reasonable length.

\subsection{Precision}
\label{sec-ideas12}

The estimator assigns the same rank $c_j$ to any password $\alpha$ for which the probability falls within the range $p(\beta_j) > p(\alpha) \geq p(\beta_{j+1})$.
Intuitively, passwords with distinct probabilities should not get the same numeric estimate. Certainly, this is not an issue when the password strength is presented on a reduced scale using descriptive characteristics like \textsl{weak} -- \textsl{medium} --\textsl{strong} -- \textsl{very strong}, or using a traffic lights metaphor \textsl{red} -- \textsl{amber} -- \textsl{green}.
\medskip

\begin{figure}[h]
\centering
\includegraphics[scale=0.65]{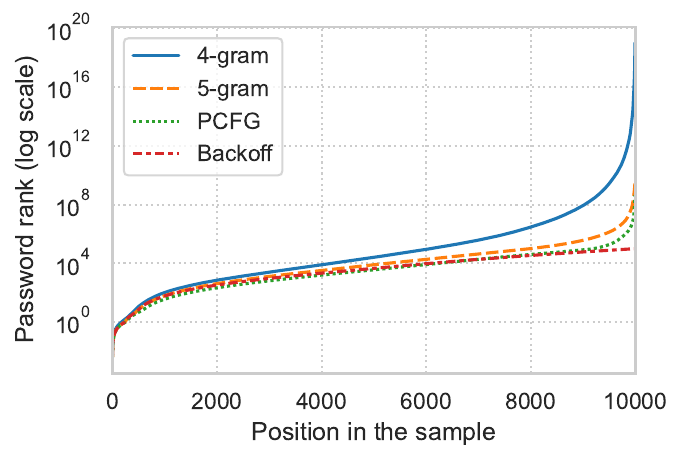}
\caption{Password ranks corresponding to the position in the sample}
\label{fig2}
\end{figure}

\begin{idea}\label{inter}
Interpolate rank values within intervals using an appropriate function. The most basic approach, without additional parameters, is linear interpolation. This has no impact on memory complexity and a negligible impact on time complexity. Figure \ref{fig2} shows a graph of password ranks, on a logarithmic scale, for various models and the sample size of 10,000. It appears that linear interpolation on the logarithmic scale should perform well for these models.
\end{idea}

The precision of the estimator depends on the sample size. More specifically, it depends on the number of unique probabilities in the set $P = \{p(\beta_1), \ldots, p(\beta_n)\}$. We define the overlap of $\Theta$ as the fraction of probability values that are already in the set, and therefore do not contribute to the estimator's precision: $1- |P|/n$. Table \ref{tab-overlap} shows the average overlap for different models and sample sizes. Surprising differences in overlap are observed among different models. An expected increase in overlap is observed with an increasing sample size, since the overlap depends substantially on password probability distribution, given by the model from which the passwords are generated. On the other hand, a larger training dataset results in greater diversity of passwords, leading to slightly lower overlap.
\medskip

\begin{idea}\label{sampl}
The estimator will sample random passwords for $\Theta$ until it gets $n$ unique probabilities. It \emph{compresses} sample by discarding duplicate probabilities in such a way that preserves the cumulative sum of the entry with the largest index. Hence, the rank calculation remains intact, and the overlap of the resulting $\Theta$ is be $0$. Since the sampling is done in precomputation phase, it does not impact the estimation time or memory complexity in any way.
\end{idea}

\begin{table}[h]\centering
\begin{tabular}{rrrrrr}\toprule
training set & sample size & 4-gram & 5-gram & Backoff & PCFG \\\midrule
500,000
   & 10,000 & 13.6\% & 16.5\% & 20.4\% & 44.6\% \\
   & 30,000 & 20.6\% & 25.8\% & 31.6\% & 60.8\% \\
   & 50,000 & 24.8\% & 30.5\% & 37.3\% & 67.1\% \\\midrule
1,000,000
   & 10,000 & 12.4\% & 14.7\% & 17.2\% & 43.1\% \\
   & 30,000 & 19.1\% & 22.4\% & 27.6\% & 58.5\% \\
   & 50,000 & 22.4\% & 26.7\% & 33.2\% & 64.8\% \\\bottomrule  
\end{tabular}
\caption{Overlap percentage for different models and sample sizes. Models are trained on 500,000 and 1,000,000 passwords using the most frequent passwords from the RockYou dataset. Every number is an average of 3 experiments.}\label{tab-overlap}
\end{table}

Table \ref{tab-compress} shows how many passwords must be sampled using a trained model to achieve the target size of the sample with distinct probabilities.

\begin{table}[h]\centering
\begin{tabular}{rrrrr}\toprule
target      & \multicolumn{4}{c}{Sampled passwords}\\
sample size & 4-gram & 5-gram & Backoff & PCFG \\\midrule
 10,000     & 11,689 & 12,239 & 12,894 &  23,483 \\
 30,000     & 38,795 & 42,032 & 47,178 & 122,865 \\
 50,000     & 68,358 & 75,953 & 90,123 & 258,489\\\bottomrule  
\end{tabular}
\caption{Average number of sampled passwords required to achieve the desired sample size with distinct probabilities. Models are trained on 500,000 passwords using the most frequent passwords from the RockYou dataset. Every number is an average of 10 experiments, rounded to the nearest integer.}\label{tab-compress}
\end{table}

\subsection{Estimation speed}
\label{sec-idea3}

The binary search employed in the original estimator is fast enough for assessing individual passwords. However, when the estimator is used to evaluate or compare different models and their variants, the ranks of a large number of passwords need to be estimated. An optimization can be relevant in these scenarios.

\begin{idea}
Divide the interval of possible probability values $p(\beta_i)$, in our case expressed as negative $\log_2$ values, into $t$ intervals (bins): $[0,\tau_1)$, $[\tau_2,\tau_3)$, \ldots, $[\tau_{t-1}, \infty)$, where $0<\tau_1 <\ldots <\tau_{t-1}$. For each interval, we calculate minimal and maximal look-up indices that narrow interval for binary search (we use $\tau_0=0$ in the following equations):
\begin{align*}
 \text{LU}_\text{min}(i) &= \max\{ 1\leq j \leq n \mid -\log_2 p(\beta_j) \geq \tau_{i-1} \}, \\ 
 \text{LU}_\text{max}(i) &= \min\{ 1\leq j \leq n \mid -\log_2 p(\beta_j) < \tau_{i-1} \}, \text{ for } 1\leq i \leq t.
\end{align*}
\end{idea}

The estimator is adapted accordingly. Given a password $\alpha$, we calculate an appropriate interval such that $-\log_2 p(\alpha) \in [\tau_{i-1}, \tau_i)$. Then, the binary search is performed within the set of indices $\{\text{LU}_\text{min}(i),\ldots , \text{LU}_\text{max}(i)\}$, instead of full set $\{1,\ldots , n\}$. We expect to narrow the interval for the binary search substantially, so the benefit of fewer comparisons will be measurable. Trivially, the precision of the estimator remains unchanged.

The price paid is the cost of computing $\text{LU}_\text{min}$ and $\text{LU}_\text{max}$ arrays, which is simple one-time precomputation, and small memory needed to store these arrays in the estimator\footnote{For example, 100 intervals ``cost'' approximately 7.8 kB, even with a wasteful representation using Python's \texttt{int} objects for stored indices and lists for the arrays}.

\section{Experiments}
\label{sec4}

We implement the ideas presented in the previous section and present the results of our experiments. 

\subsection{Precision}

The ideas aimed at improving precision apply to the Monte Carlo Estimator, regardless of the underlying model. We do not attempt to modify the models. For example, if password \textsl{-1-1-1-1} is assigned \texttt{inf}\footnote{Python's \texttt{float(`inf')} value} as negative $\log_2$ probability in the PCFG model, because the pattern is outside of the trained grammar, we do not try to ``fix this''. Moreover, we do not compare the performance of the models to each other.

We assess the impact of our ideas on the real ranks of password generated by the models. Similarly to the original paper \cite{AM15}, we generate all passwords up to some probability threshold. The rank of a password is its position in the list sorted by the probabilities assigned by the model to the passwords. 

The first experiment uses the PCFG model trained on 10 million passwords from the RockYou dataset. The threshold for password generation was set at $20$, i.e., all passwords with probability at least $2^{-20}$ were generated -- there were 91,693 passwords in this dataset (let's denote it $T$). We consider various combinations of proposed ideas:

\begin{itemize}
\item original -- a reference implementation of the estimator \cite{A16};
\item interpolation -- interpolate rank calculation within the interval between two adjacent probabilities (Idea \ref{inter});
\item sampling -- improved sampling with $n$ unique probabilities (Idea \ref{sampl});
\item all -- a combination of interpolation and sampling.
\end{itemize}

Let $\text{rr}(\alpha)$ denote the real rank of password $\alpha\in T$, and let $\text{er}(\alpha)$ denote the rank estimated by a particular variant of the estimator. The \textsl{weighted error} of the estimator on the password set $T$ is calculated as follows:
\[
  \sum_{\alpha\in T} p(\alpha)\, |\text{er}(\alpha) - \text{rr}(\alpha)|.
\] 

\noindent The weighted error assumes that the estimators are used to asses passwords chosen by humans, following the original distribution. We also consider a \textsl{simple error} for completeness:
\[
  \frac{1}{|T|}\, \sum_{\alpha\in T} |\text{er}(\alpha) - \text{rr}(\alpha)|.
\] 

\begin{table}[h]\centering
\begin{tabular}{lrr}\toprule
variant       & weighted error & simple error \\\midrule
original      & 16.54 & 101.11 \\
interpolation & 15.33 & 90.63 \\
sampling      & 11.79 & 70.63 \\ 
all           & 10.86 & 63.10 \\\bottomrule  
\end{tabular}
\caption{Weighted and simple errors of various estimator variants. Every number is an average of 100 experiments.}\label{tab-experiment}
\end{table}

Table \ref{tab-experiment} shows the results of our experiment. We performed 100 experiments. We have to warn the reader -- the reported errors are sensitive to the particular password distribution sampled into $\Theta$. Unsurprisingly, the sampling (Idea \ref{sampl}) helps to reduce estimation errors in general. The situation with interpolation (Idea \ref{inter}) is mixed, with a substantial fraction of experiments showing worse statistics. The reason is that the interpolation makes the error worse when passwords in $\Theta$ already ``overshoot'' their true ranks. Taking the same rank without interpolation compensates for this. Therefore, interpolation cannot be recommended for improving the precision of the estimator. On the other hand, it helps with the ``same rank'' problem, when different passwords are assigned the same rank by the estimator.

\begin{figure}[h]
\centering
\includegraphics[scale=0.65]{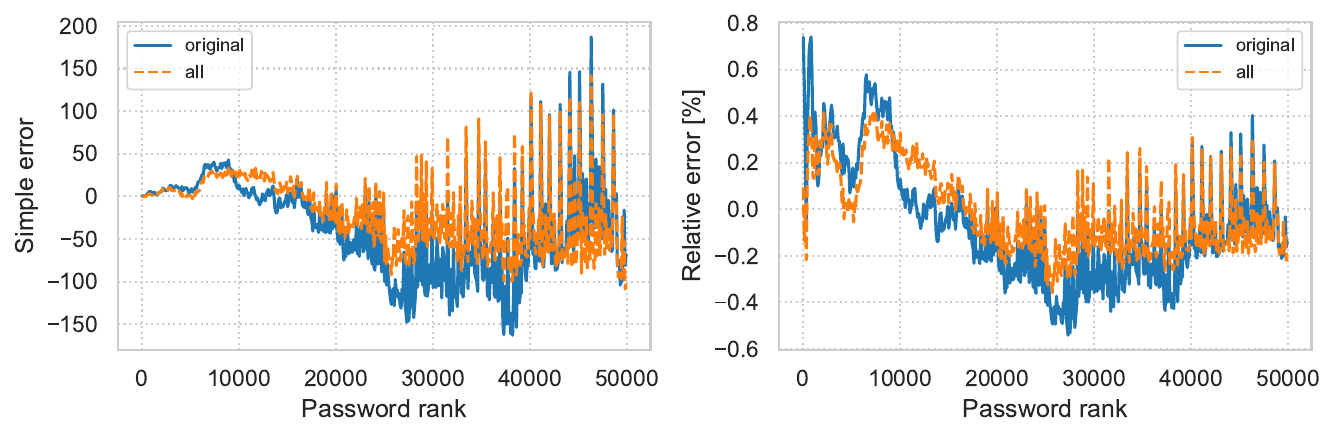}
\caption{The simple difference error (on the left) and the relative error (on the right) of the original and the "all" strategy. Both graphs display results for 50,000 of the most probable passwords from the PCFG model. The numbers are the average values from 100 experiments.}
\label{fig3}
\end{figure}

Figure \ref{fig3} compares visually the original variant with the ``all'' variant. It illustrates the simple difference of calculated rank and estimated rank. It also shows the relative error of the estimators. As expected, based on the convergence proof in \cite{AM15}, the relative error is rather small in both cases. 

\subsection{Estimation speed}

We tested two configurations: the first one with 100 intervals (bins), and the second with 1,000 intervals. Negative log probabilities are divided into fixed intervals $[0,1)$, $[1,2)$, \ldots, $[99,\infty)$ for the first case, and into $[0,0.1)$, $[0.1,0.2)$, \ldots, $[99.9,\infty)$ for the second case, respectively. Both configurations were tested with four different sizes of $\Theta$. Table \ref{tab-estspeed} shows the relative speed of different variants with respect to the baseline, which is the original algorithm with $|\Theta| = 10000$. The results confirm a moderate speed-up for 100 intervals and a substantial speed-up for 1,000 intervals.

\begin{table}[h]\centering
\begin{tabular}{rrrr}\toprule
            & \multicolumn{3}{c}{Estimation performance}\\
sample size & original & 100 bins & 1000 bins \\\midrule
 10,000     & 1.00     & 0.92     & 0.37 \\
 30,000     & 1.08     & 1.00     & 0.39 \\
 50,000     & 1.13     & 1.04     & 0.40 \\
 100,000    & 1.20     & 1.10     & 0.40 \\\bottomrule  
\end{tabular}
\caption{Average relative estimation performance, where the baseline $1.00$ is the estimation performance of the original binary search for the sample size 10,000. Experiment uses $10^6$ randomly generated passwords by the PCFG model. Every number is an average of 10 experiments, and rounded to the two decimal places.}\label{tab-estspeed}
\end{table}

\section{Additional observation and conclusion}

\begin{figure}[h]
\centering
\includegraphics[scale=0.65]{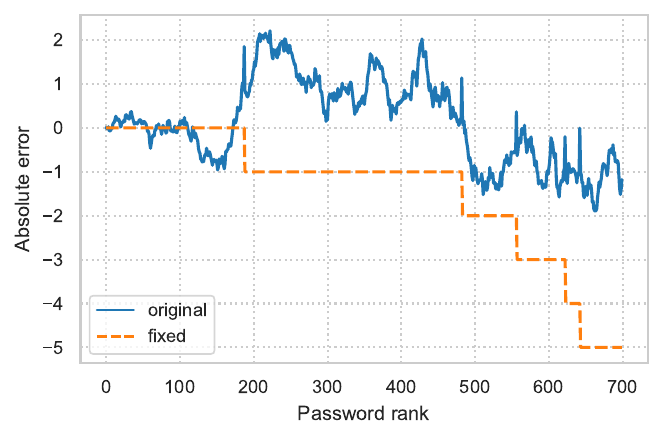}
\caption{Comparison of original password's rank estimate and estimate according the position in sampled passwords (denoted as \emph{fixed}).}
\label{fig4}
\end{figure}

Since passwords in $\Theta$ are generated according to their probability, with sufficiently large sample size, we expect that for some $k$, the top-$k$ most probable passwords will be in the correct order at the beginning of $\Theta$. Therefore, simply reporting the order of these top-$k$ passwords by the estimator can be beneficial to the precision. Figure \ref{fig4} illustrates this phenomenon for the PCFG model and the sample size of 10,000, where approximately the top 180 passwords have the exact rank as their position in $\Theta$. However, further down the precision quickly deteriorates. 

An interesting question is if we can improve the estimator's precision by compensating for unusually large or small jumps (differences) between adjacent 
probabilities in the sampled passwords.

An area outside this paper that deserves further focus is the precision of the estimator for low-probability passwords. The estimator's precision worsens for passwords with high ranks. A potential approach might use a different or additional sampling methods that focus on less probable passwords, so that we can cover this part of the probability space better.

\end{document}